\documentclass[10pt,twocolumn]{article} % 两栏排版；若想单栏，去掉 twocolumn
\usepackage[margin=1in]{geometry}       % 页边距，可按需调整
\usepackage{authblk}                     % 作者/单位
\usepackage{amsmath,amssymb}
\usepackage{graphicx}
\usepackage{hyperref}
\usepackage{xcolor}
\usepackage{url}
\usepackage{newtxtext,newtxmath}         % Times 风格字体（可选）
\newtheorem{proposition}{Proposition}%% lep
\usepackage[numbers,sort&compress]{natbib}
\providecommand{\keywords}[1]{\noindent\textbf{Keywords: } #1}

\title{Radial spoke energy for self-navigated motion detection and position-ordered dynamic musculoskeletal MRI}

\author[1]{Enping Lin\thanks{Corresponding author: \href{mailto:enping.lin@childrens.harvard.edu}{enping.lin@childrens.harvard.edu}}}
\author[1]{Fatih Calakli}
\author[1]{Musa Tun\c{c} Arslan}
\author[1]{Giovani Schulte Farina}
\author[1]{Simon Keith Warfield}

\affil[1]{Computational Radiology Laboratory, Department of Radiology,
Boston Children's Hospital, Harvard Medical School, Harvard University,
401 Park Drive, Boston, MA 02115, USA}

\date{} % arXiv 通常不显示日期

\begin{document}
\maketitle

% ===== 摘要 =====
\begin{abstract}
Motion remains a key challenge and research focus in MRI, as both involuntary
(e.g., head movement) and voluntary (e.g., joint motion) motion can degrade
image quality or offer opportunities for dynamic assessment. Existing motion
sensing methods—like external tracking or navigator sequences—often require
extra hardware, increase SAR, or demand sequence modification, limiting
clinical flexibility. We propose a computationally efficient, self-navigated
motion sensing technique based on spoke energy derived from 3D radial k-space
data. Using the Fourier Slice and Parseval’s Theorems, spoke energy captures
object-coil alignment and can be computed without altering the sequence. A
sliding window summation improves robustness, and a 2nd principal
component-based (2ndPCA) strategy yields a unified motion-sensitive signal.
Beyond conventional head motion correction, we demonstrate the method’s novel
application in enhancing dynamic 4D MRI of the ankle and knee under the
continuous movement protocol. By sorting spokes based on position rather than
time, we achieve motion-resolved reconstructions with improved anatomical
clarity. This method enables real-time motion detection and supports broader
adoption of motion-aware, dynamic MRI.
\end{abstract}


% ===== 正文从这里开始 =====
% \section{Introduction}
% ...

%% ========================= main text ====================================

\section{Introduction}
\label{sec1}
Magnetic Resonance Imaging (MRI) is an essential tool in clinical diagnostics, renowned for its ability to provide high-resolution, soft-tissue contrast images without ionizing radiation. Its non-invasive nature makes it highly suitable for evaluating a wide range of anatomical and physiological conditions, including those in the brain, heart, and musculoskeletal system.

Motion has emerged as a critical aspect of MRI, presenting both challenges and opportunities. On one hand, involuntary or unintentional patient motion—such as head movement, breathing, or subtle shifts—can introduce inconsistencies in the k-space data, leading to artifacts such as blurring or ghosting\citep{Vakli2023}. These motion-induced artifacts may significantly degrade image quality, obscure anatomical detail, and reduce diagnostic accuracy. For populations such as pediatric patients or individuals with neurological or cognitive impairment, this often necessitates sedation, repeat scans, or even scan termination, thereby increasing scan time, cost\citep{Slipsager2020,Afacan2016,Andre2015,Vanderby2010}, and clinical risk\citep{Malviya2000,Havidich2016}.

On the other hand, voluntary or physiological motion during MRI is often the target of measurement itself. Dynamic imaging applications—such as cardiac cine MRI\citep{Pan2024}, joint motion studies (e.g., knee flexion/extension)\citep{Huang2022}, and functional assessments—rely on accurately capturing and resolving motion across time. In these contexts, motion is not a nuisance to be corrected, but rather a signal to be measured, tracked, and reconstructed with high temporal fidelity. Therefore, developing reliable motion sensing technologies is essential not only for correcting undesired motion artifacts but also for enhancing dynamic MRI applications where motion is intentional and meaningful.

In recent years, real-time motion sensing has become increasingly important in both domains. For involuntary motion, real-time feedback allows scan operators to take corrective action—either by pausing, re-scanning, or applying prospective or retrospective motion correction algorithms. For dynamic imaging, real-time motion tracking provides the temporal markers needed to segment and reconstruct motion-resolved frames, supporting more accurate and physiologically meaningful reconstructions\citep{Coll-Font2021}.

Various techniques have been proposed for real-time motion sensing in MRI. One class of approaches uses external tracking systems, such as Tracoline (TCL) optical sensors\citep{Tracoline}, electromagnetic trackers\citep{Wallace2019,Afacan2020}, Moiré phase tracking (MPT), or the Kineticor system\citep{kinetic}. While effective in capturing rigid-body head or body motion, these systems require additional hardware, setup time, and often patient compliance with wearing physical markers. Moreover, they may be limited in detecting internal or non-rigid motion.

Another major category involves navigator-based techniques, such as FIDnav\citep{FIDNav,Tess2024,Wallace2020a,Wallace2020b,Wallace2020c}, FatNav\citep{FatNav}, and other navigators\citep{Ulrich2024,Welch2002,Tisdall2012}. These approaches embed motion-sensitive readouts within the imaging sequence itself, offering direct motion information without external hardware. However, they often require modifications to the original pulse sequence, introduce additional scan time, and may increase the Specific Absorption Rate (SAR), which is a limiting factor in some patient populations.

To address the limitations of existing motion sensing approaches, this work introduces a novel method based on spoke energy in 3D radial MRI. Compared to conventional Cartesian sampling, 3D radial trajectories are inherently more robust to motion and well-suited for retrospective motion correction, offering advantages for both motion measurement and correction tasks\cite{Anderson2014,Piccini2011,Solomon2021}. In 3D radial MRI, spoke energy is derived directly from k-space data and reflects the spatial distribution of the anatomy relative to the coil sensitivities. To enhance motion sensitivity while preserving high temporal resolution, we compute the summed spoke energy using a sliding window, rather than relying on individual spoke measurements. Spoke energy can be computed directly from k-space data without requiring additional hardware or navigator sequences, and its lightweight computation enables rapid motion detection in real time. Motion can thus be detected by tracking energy variations, as motion alters the spatial relationship between anatomy and fixed coils, thereby changing signal energy.

This study demonstrates that spoke energy is a simple, efficient, and effective self-navigated metric for motion sensing in 3D radial MRI, and illustrates its application in enhancing 4D (3D images + temporal) dynamic musculoskeletal imaging, which was recently presented by our group\citep{Giovani2024,Fairhurst2025}, under the novel continuous movement protocol. The major contributions are summarized as follows:

\begin{enumerate}
\item We propose a self-navigated motion sensing method based on spoke energy, requiring no external tracking devices or additional navigator sequences. The approach is computationally lightweight and supports real-time motion detection. Furthermore, the underlying principles of spoke energy in motion sensing are thoroughly explained.

\item A window-summed energy strategy is introduced to improve motion detection robustness. This method enables spoke-level detection of motion onset and offset, and the window-summed strategy can be employed in other motion metrics, such as the Center of Mass (CoM). 

\item We present a PCA-based multi-coil combination method, using the second principal component to integrate motion-sensitive variations across coils more effectively than the first component. This enhances signal clarity and facilitates intuitive motion detection and position representation.

\item Beyond its established application in head motion correction, we extend the use of spoke energy to a novel domain—serving as a positional metric in  4D dynamic musculoskeletal imaging with the continuous movement pattern. By sorting radial spokes based on their spoke energy and applying a sliding-window reconstruction strategy, we generate motion-resolved 3D dynamic musculoskeletal image series (4D cine). This approach enables dynamic volume visualization of joint motion and shows strong potential for identifying and evaluating joint impingement or pain through time-resolved anatomical assessment.
\end{enumerate}

\section{Methods}
\subsection{3D Radial MRI}

In 3D Radial MRI, data are sampled in a non-Cartesian manner by acquiring radial spokes in the 3D k-space, which radiate from the center outward. Each spoke is defined by specific k-space coordinates based on spherical angles \( \theta \) and \( \phi \). These coordinates can be expressed by the following equations:

For the x-component:
\begin{equation}
k_x = k \sin(\theta) \cos(\phi)
\end{equation}

For the y-component:
\begin{equation}
k_y = k \sin(\theta) \sin(\phi)
\end{equation}

For the z-component:
\begin{equation}
k_z = k \cos(\theta)
\end{equation}

Here, \( k \) represents the radial distance. A sufficient number of spokes ensures uniform coverage of k-space, typically achieved using techniques like the Golden-Angle or Spiral Phyllotaxis, which optimize sampling efficiency. 3D Radial sampling offers several advantages over traditional Cartesian sampling. Its motion robustness stems from the fact that each spoke passes through the center of k-space, capturing critical low-frequency information repeatedly. This redundancy makes it less susceptible to artifacts from sudden patient movements and provides a more stable basis for retrospective motion correction. Additionally, radial trajectory is convenient for retrospective motion correction. 

\subsection{Spoke Energy}

In 3D Radial MRI, we assume that \( N \) radial spokes are collected across \( M \) receiver coils. Let the vector \( \mathbf{k}_{ij} \) (where \( i = 1, 2, \dots, N \) and \( j = 1, 2, \dots, M \)) represent the \( i \)-th (in temporal order) k-space radial spoke acquired by the \( j \)-th coil.

The energy of the \( i \)-th radial spoke for the \( j \)-th coil can be calculated as:
\begin{equation}
E_{ij} = \sum_k |\mathbf{k}_{ij}[k]|^2
\end{equation}
where \(k\) represents the discrete frequency components of the k-space spoke.

For each given coil \( j \), the spoke energy \( E_{ij} \) across all spokes \( i \) can serve as a metric to monitor motion. As motion occurs, changes in the subject’s anatomy alter the spoke energy, providing a real-time indication of movement. By monitoring variations \( E_{ij} \) over time (i.e., across successive spokes), we can detect head motion or other involuntary subject movements. Significant fluctuations in \( E_{ij} \) along spokes indicate shifts in the spatial configuration of the anatomy relative to the coils, making spoke energy an effective metric for motion sensing.

Moreover, the calculation of spoke energy is computationally efficient and can be performed on-the-fly. This approach enables real-time motion sensing, providing MRI practitioners with crucial information to discard and re-scan shots affected by excessive motion, as well as to guide retrospective motion correction.

\subsection{How Spoke Energy Enables Motion Sensing}

According to the Fourier Slice Theorem, the Inverse Fourier Transform of a k-space radial spoke corresponds to a one-dimensional projection of the image volume along the direction of that spoke. When the subject remains still, the energy of these one-dimensional projections should stay relatively constant, regardless of direction. However, if the subject moves, the anatomy within the imaging volume shifts accordingly. Since the receiver coils are fixed, this movement causes the anatomy to correspond to different coil sensitivities. As a result, the received intensity energy of the one-dimensional projections varies with the subject's motion. This sensitivity variation allows us to utilize the intensity energy of these projections for motion sensing.

 Parseval’s Theorem indicates that the energy of a k-space radial spoke is equal to the energy of its corresponding projection in the image space. Therefore, calculating the energy for each k-space spoke effectively captures changes in the spatial configuration of the anatomy. The effectiveness of spoke energy as a metric for motion sensing arises from the non-uniform distribution of coil sensitivity across the field of view.

\subsection{Sum Spoke Energy Using a Sliding Window}

Even when a subject remains still within a shot, in MPnRAGE\citep{MPnRAGE}, the energy of individual spokes can vary due to the relaxation effects of tissues. To enhance motion-sensing sensitivity, it is beneficial to use the summed spoke energy over a sliding window with a length that is an integer multiple of the shot duration. This approach helps to offset relaxation-derived variations in spoke energy.

Assume that the total number of spokes is \(N\), and the sliding window has a length of \(L\) and a stride of \(S\). The summed spoke energy in the \(n\)-th window for the \(j\)-th coil is denoted as \(W_{nj}\), and is defined as:

\begin{equation}
W_{nj} = \sum_{i=(n-1)\times S+1}^{(n-1)\times S+L} E_{ij}, \quad \text{s.t.} \quad
\begin{cases}
(n-1)\times S + L \leq N \\
n \geq 1
\end{cases}
\end{equation}

Another key advantage of summed spoke energy over individual spoke energy is its robustness to noise perturbations. Individual spoke energy is more susceptible to noise, which can compromise accurate motion sensing. Summing spoke energy across multiple spokes increases the signal-to-noise ratio (SNR), resulting in a more robust motion-sensing metric. In motion-detection applications, two critical questions arise:
\begin{enumerate}
\item The temporal resolution of motion detection: If a subject moves, what is the minimum accuracy with which we can determine the start and end of the motion event along the temporal axis?    
\item The temporal resolution of stillness detection: If a subject remains still for a certain duration, what is the minimum period required to recognize the subject as being still?
\end{enumerate}

The first question depends on the stride of the sliding window. If the stride of the sliding window is \(S\) spokes, the temporal resolution for motion detection is the time required to collect \(S\) spokes. This means that motion can only be detected within a span of \(S\) spokes, limiting the precision of identifying the exact start or end time of the motion event. Consequently, the stride cannot be too large. Throughout this work, we set \(S=1\) for highly temporal motion detection. The second question depends on the length of the sliding window. If the sliding window length is \(L\) spokes, the temporal resolution for stillness recognition is the time required to collect \(L\) spokes. This implies that if the subject remains still for a duration shorter than the collection time of \(L\) spokes, the still state will not be detected. Therefore, the window length must also be carefully chosen to avoid being too large. 

\begin{figure*}%% placement specifier
\centering%% For centre alignment of image.
\includegraphics[width=\linewidth]{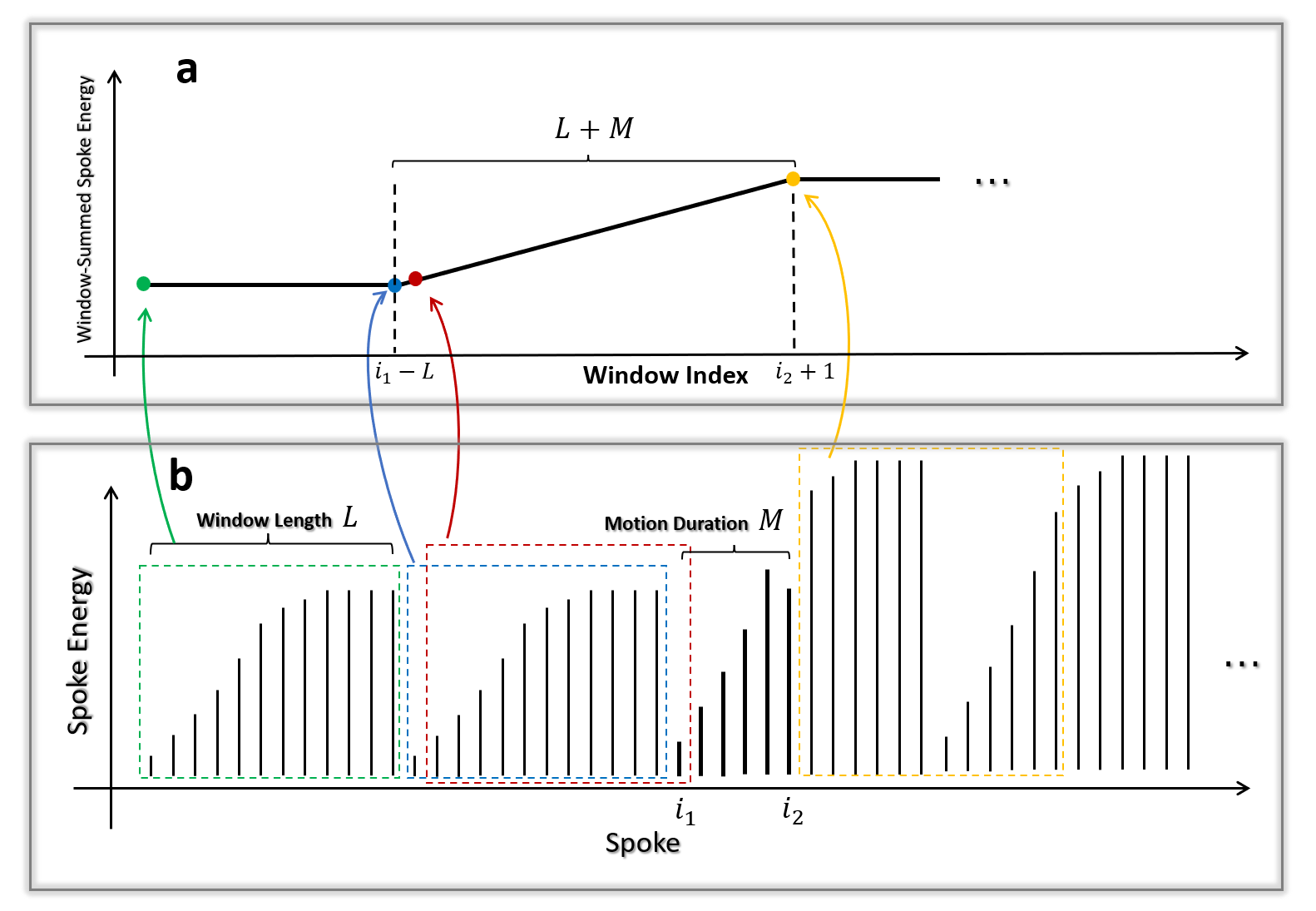}
%% Use \caption command for figure caption and label.
\caption{Illustration of the relationship between sliding-window-summed spoke energy (a) and individual spoke energy (b). In panel (b), the spoke energy during the motion duration \(M\) is highlighted in bold, which starts at \(i_1\) and ends at \(i_2\), i.e., \(i_2-i_1=M-1\). For demonstration purposes, four windows are marked with dashed rectangles of different colors (green, blue, red, and yellow). The green dashed rectangle represents the first window, the blue rectangle indicates the last window before the motion event, the red rectangle marks the first window encompassing motion spokes, and the yellow rectangle corresponds to the first window after the motion event. Their respective window-summed spoke energies are indicated on the curve in panel (a) using the same colors. The window length \(L\) equals the number of spokes in a shot. Within a shot, spoke energy may vary in a specific pattern due to relaxation effects and other factors, even when the subject is stationary. During a motion event lasting \(M\), the window-summed spoke energy is affected for a duration of \(L+M\) (represented by the interval between the blue and yellow dots). This relationship demonstrates the interplay between motion events and the dynamics of window-summed spoke energy.} \label{fig_SE} 
\end{figure*}

Another important question is the relationship between the start and end of variations in the window-summed spoke energy and the corresponding start and end of motion events. To address this, we provide a clear illustration in Figure \ref{fig_SE} and state in Proposition \ref{prop1}.

\begin{proposition} \label{prop1}
 If a motion event begins at the \(i_1\)-th spoke and ends at the \(i_2\)-th spoke (\(L<i_1\leq i_2 < N-L\)), with the motion event duration denoted as \(M=i_2-i_1+1\), the variation in the window-summed spoke energy \(W_{nj}\) for the \(j\)-th coil starts at \(i_1-L\) and ends at \(i_2+1\), resulting in a total variation duration of \(M+L\). Conversely, if the variation in the window-summed spoke energy is observed to start at \(j_1\) and end at \(j_2\), it can be inferred that the motion event begins at the \((j_1+L)\)-th spoke and ends at the \(max(j_2-1,j_1+L)\)-th spoke.
\end{proposition}

It is noteworthy that, apart from spoke energy, other metrics, e.g., CoM\citep{anderson2013,Calakli2024ISMRM} and low-resolution scout images\citep{kecskemeti2018}, could also be used in this sliding window approach for a robust motion sensing. However, spoke energy is advantageous for on-the-fly motion monitor due to its significantly simple computation.

\subsection{Combine Multi-Coil Sliding Window Spoke Energy Signals using 2ndPCA}

Theoretically, the Sliding Window Spoke Energy Signals from each coil should reflect motion-induced variations. However, due to the spatially varying coil profiles, each coil exhibits different sensitivities to various motion states. This inherent variability means that certain coils may be more sensitive to specific types of motion than others. By leveraging the signals from multiple coils, the robustness of motion sensing can be significantly enhanced.

While retaining multi-coils signals improves motion detection, it complicates the direct observation of motion-induced variations, making the results less intuitive. To address this, we combine the multi-coil sliding window spoke energy signals into a single representative signal in this study. This combination is achieved using Principal Component Analysis (PCA), which allows us to summarize the key features of multi-coil signals.

Rather than using the first principal component, we deliberately select the second principal component as the combined signal. The first component often captures the strongest but relatively constant features, which fail to reflect motion-induced variations across coils. In contrast, the second component retains strong but dynamic features that are more sensitive to motion. Higher-order components tend to be dominated by noise and contribute little to motion detection. We refer to this strategy as 2ndPCA for clarity and will discuss more in \ref{sec_PCA}. The resulting single, scalar-valued signal is intuitive for identifying motion events and can serve as a proxy for object position over time.

\section{Experiments and Material}
All numerical simulations and reconstructions were conducted on a high-performance computing server running Ubuntu with dual AMD EPYC 7542 32-core processors (64 threads total) and 1 TB of RAM. MATLAB R2023b was used as the primary programming environment. Our code and synthetic data can be available by a reasonable request.

\subsection{Simulated Experiment}

\begin{figure*}%% placement specifier
\centering%% For centre alignment of image.
\includegraphics[width=\linewidth]{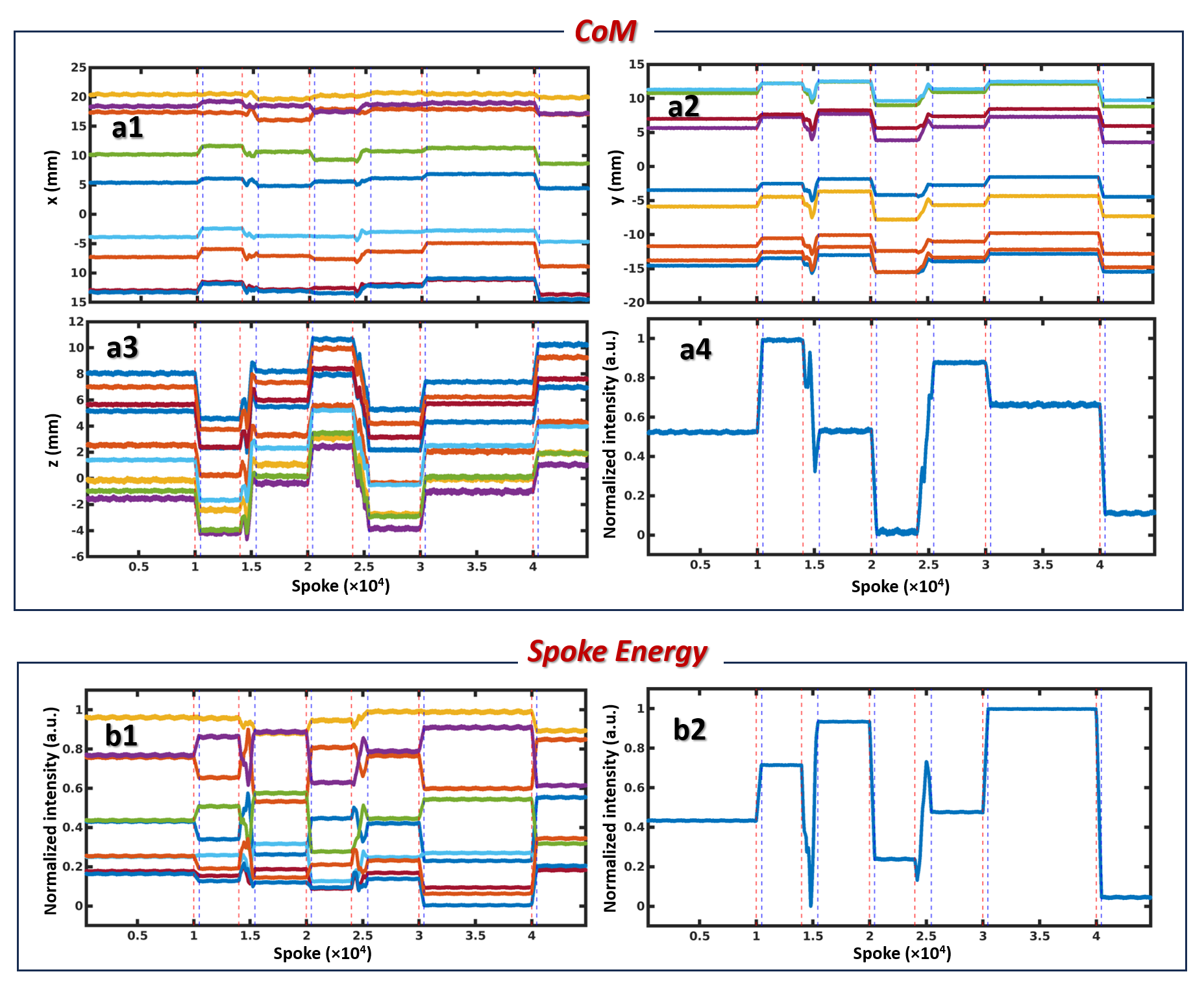}
%% Use \caption command for figure caption and label.
\caption{Motion sensing of simulated head motion data using a sliding shot-length window. Six motion events are injected in the simulated data, which begin at spoke indices [10000,14000,20000,24000,30000,40000] and end at [10000, 15000, 20000, 25000, 30000, 40000], respectively. (a1–a3) display the curves of the CoM coordinates in the x, y, and z directions, respectively, with different colors representing individual coil signals. (a4) shows a composite curve generated by combining all CoM signals using 2ndPCA in (a1–a3) to facilitate motion tracking. The CoM is calculated from the spokes within the sliding window using Least-Squares minimization, with a total computation time of 1 hour and 34 minutes. (b1) presents the sliding window-sum spoke energy curves, and (b2) shows a 2ndPCA combined curve from all signals in (b1) for simplified motion tracking. The total calculation time for sliding window-sum spoke energy is under 1 second. Red and blue dashed vertical lines indicate the ideal starting [9552, 13552, 19552, 23552, 29552, 39552] and ending [10001, 15001, 20001, 25001, 30001, 40001] of the motion events in the axis of sliding window index, respectively.} \label{fig_MT} 
\end{figure*}

In Figure \ref{fig_MT}, we synthetically injected six motion events in a still 3D head MPnRAGE data, and utilized CoM and spoke energy, derived from a sliding window, to track the motion. The six motion events begin at spoke indices [10000,14000,20000,24000,30000,40000] and end at [10000, 15000, 20000, 25000, 30000, 40000], respectively. The CoM is calculated from the spokes within each sliding window using Least-Squares minimization. 2ndPCA is used to combine CoM signal and spoke energy signal, respectively. This simulated experiment is aimed to verify the Proposition \ref{prop1} and make a comparison between CoM and Spoke energy. The result is shown in Figure \ref{fig_MT}.

\subsection{Head Motion Correction Experiment}

To validate the proposed approach in practical head motion detection, two volunteers were scanned at 3T using a MAGNETOM Prisma (Siemens Healthcare, Erlangen, Germany) with a 64-channel head coils. An MPnRAGE sequence was used to acquire a pseudo-random 3D golden angle radial sampling pattern to generate a set of spatially and temporally uniformly distributed spokes. Volunteers provided written informed consent, and all scans adhered to the local review board protocol. Each volunteer underwent two scans: one while holding still and another with instructed head repositioning. $T_{1}$-weighted structural images with 0.82 mm isotropic resolution were acquired from Volunteer 1 which is referred to as Head Data 1, and with 0.875 mm isotropic resolution from Volunteer 2 which is referred to as Head Data 2. Volunteer 1 repositioned their head every 60 seconds, while Volunteer 2 moved their heads randomly after the second minute. The total acquisition time for each scan was 6.4 minutes, with 44,800 spokes. Radial data are received in shots. Each shot contains 448 spokes for Head Data 1 and 224 spokes for Head Data 2, respectively.

We utilize the proposed spoke energy metric to monitor the head motion and perform registration-based motion correction using flexible-length spoke subsets derived from the motion detection. The results of motion correction using these flexible-length spoke subsets are compared to those obtained with fixed-length shot subsets (10 and 20 shots). During registration-based motion correction, images corresponding to the shot subsets are reconstructed using a custom NUFFT (Non-Uniform Fast Fourier Transform) algorithm, and image registration is conducted with MATLAB's built-in image registration toolbox to obtain the six rigid-body motion parameters. Based on the motion parameters derived for each subset, motion is compensated by rotating the k-space trajectory and applying phase ramping in the k-space data. Motion-corrected images are evaluated both visually and quantitatively. For visual assessment, all motion-corrected images are compared to still-scan images, with specific structural details highlighted for clarity. For quantitative assessment, motion-corrected images are evaluated using Spectral Entropy (SE, where lower values indicate more randomness), Average Edge Strength (AES, where higher values indicate sharper edges), and Structural Similarity Index Measure (SSIM, ranging from 0 to 1, with higher values indicating greater structural similarity to no-motion images). In general, higher SE, AES, and SSIM values indicate better motion correction. The processing results for Head Data 1 and 2 are presented in Figures \ref{fig_H1} and \ref{fig_H2}, respectively. 

\subsection{Stepwise Ankle Movement Experiment}

 Stepwise dynamic imaging of the ankle joints was conducted using a 3T MAGNETOM Prisma scanner (Siemens Healthcare, Erlangen, Germany) with an 18-channel U-shaped body coil. The imaging sequence employed was a $T_{1}$-weighted SOAR/MPnRAGE\citep{MPnRAGE} protocol, with parameters set to a 0.5 mm isotropic resolution and a total of 44,800 spokes over 100 shots, with a base resolution of 384. The total scan time was approximately 6 minutes and 20 seconds. The volunteers gave their informed written consent before imaging and all scans were performed according to the local institutional review board protocol. The volunteer was instructed to move the foot to the plantar flexion position. During the movement, the volunteer was instructed to hold each position for one minute, followed by a 20-second movement, and was allowed to move the back if the scan was still ongoing.
 
\begin{figure*}%% placement specifier
\centering%% For centre alignment of image.
\includegraphics[width=\linewidth]{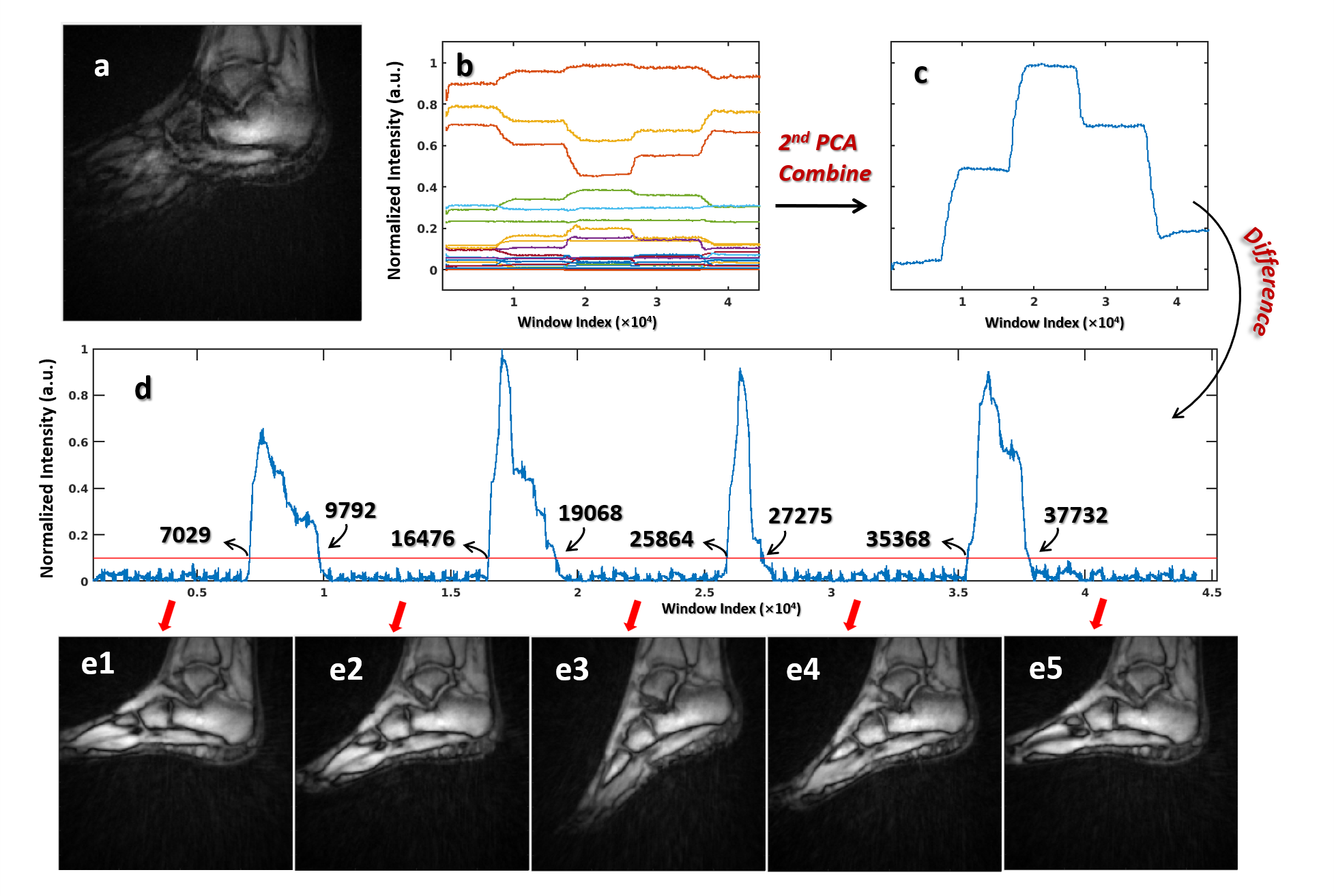}
\caption{Analysis of stepwise moving ankle data using spoke energy. (a) Reconstructed image from all 44,800 spokes. (b) Sliding window-summed spoke energy with a window length of 448 spokes, where signals from all 18 coils are represented in different colors. (c) The normalized 2ndPCA combined signal derived from the signals in panel (b) for intuitive motion observation. (d) The difference of the normalized 2ndPCA combined signal, used for motion detection. Peaks in this plot indicate significant motion events, separating the data into 5 stationary states divided by 4 peaks. The signal intensity has been normalized. A threshold of 0.1 is applied to determine the starting and ending indices of the peak bandwidths, identified as [7029, 16476, 25864, 35368] and [9792, 19068, 27275, 37732], respectively, which are used to determine the spoke indices for each stationary state. (e1–e5) Reconstructed images corresponding to the 5 stationary states using the identified spoke subsets according to proposition \ref{prop1}, with starting and ending spoke indices [1, 9791, 19067, 27274, 37731] and [7477, 16924, 26312, 35816, 44800], respectively.} \label{fig_A} 
\end{figure*}

 Guided by spoke energy, we can identify the stationary and motion states in segmented spoke data and reconstruct the frames corresponding to the stationary state of ankle from segmented spoke subsets. The results of this experiment are shown in Figure \ref{fig_A}. This 4D dynamic musculoskeletal imaging protocol is recently reported by our group in conferences \citep{Giovani2024,Fairhurst2025} and is promising for identify and analyzing the causes of joint impingement or pain by visualizing 3D volume of musculoskeletal structure with movement.

\subsection{Continuous Ankle and Knee Movement Experiment}

In addition to the stepwise movement protocol in our recent conference reports \citep{Giovani2024,Fairhurst2025}, we conducted a novel experiment involving continuous ankle and knee movement. In this setting, volunteers were instructed to perform repeated flexion and extension of the joints in a comfortable, continuous and cyclic pattern. Compared to the stepwise protocol, continuous motion offers the advantage of more evenly distributing spoke data across positions, which is beneficial for position-resolved dynamic imaging. However, this pattern introduces motion artifacts in conventional sliding-window reconstruction, as each window may encompass substantial movement.

To address this, we treated spoke energy as a proxy for the object’s position in the continuous motion setting. Based on the combined spoke energy signal, spoke data were reordered to integrate multiple motion cycles into a single representative period. Subsequently, a sliding-window reconstruction was applied to the position-sorted data, resulting in a high-quality 4D video with reduced motion artifacts and improved anatomical clarity. The results are shown in Figure \ref{fig_CA} and \ref{fig_CK}, and the dynamic images with 3 planes are given in Supplementary Material.

\section{Results and Discussions}

\subsection{Simulated Motion Sensing} 
As shown in Figure \ref{fig_MT}, all signals vary conspicuously in alignment with motion events, where red dashed lines indicate the starting indices of motion [9552, 13552, 19552, 23552, 29552, 39552] and blue dashed lines indicate the ending indices [10001, 15001, 20001, 25001, 30001, 40001]. These indices are obtained using ideal motion event spoke indices and Proposition \ref{prop1}. Signals remain stable when the subject is still, in both the sliding window-based CoM (Panels a1–a3) and spoke energy (Panel b1). For an intuitive visualization of motion tracking, all CoM coordinate signals (Panels a1–a3) and spoke energy signals (Panel b1) are combined into single signals in Panels a4 and b2, respectively. This experiment supports the effectiveness of sliding window-based methods in identifying motion events and demonstrates that spoke energy is comparable to CoM for this purpose. However, due to its simpler computational requirements, spoke energy is far more efficient than CoM. In this experiment, spoke energy calculation takes less than 1 second, while CoM calculation takes 1 hour and 34 minutes due to the extensive number of Least-Squares inversions required. Owing to its computational efficiency, spoke energy is well-suited for real-time motion monitoring during acquisition.

\subsection{Head Motion Correction}

\begin{figure*}%% placement specifier
\centering%% For centre alignment of image.
\includegraphics[width=\linewidth]{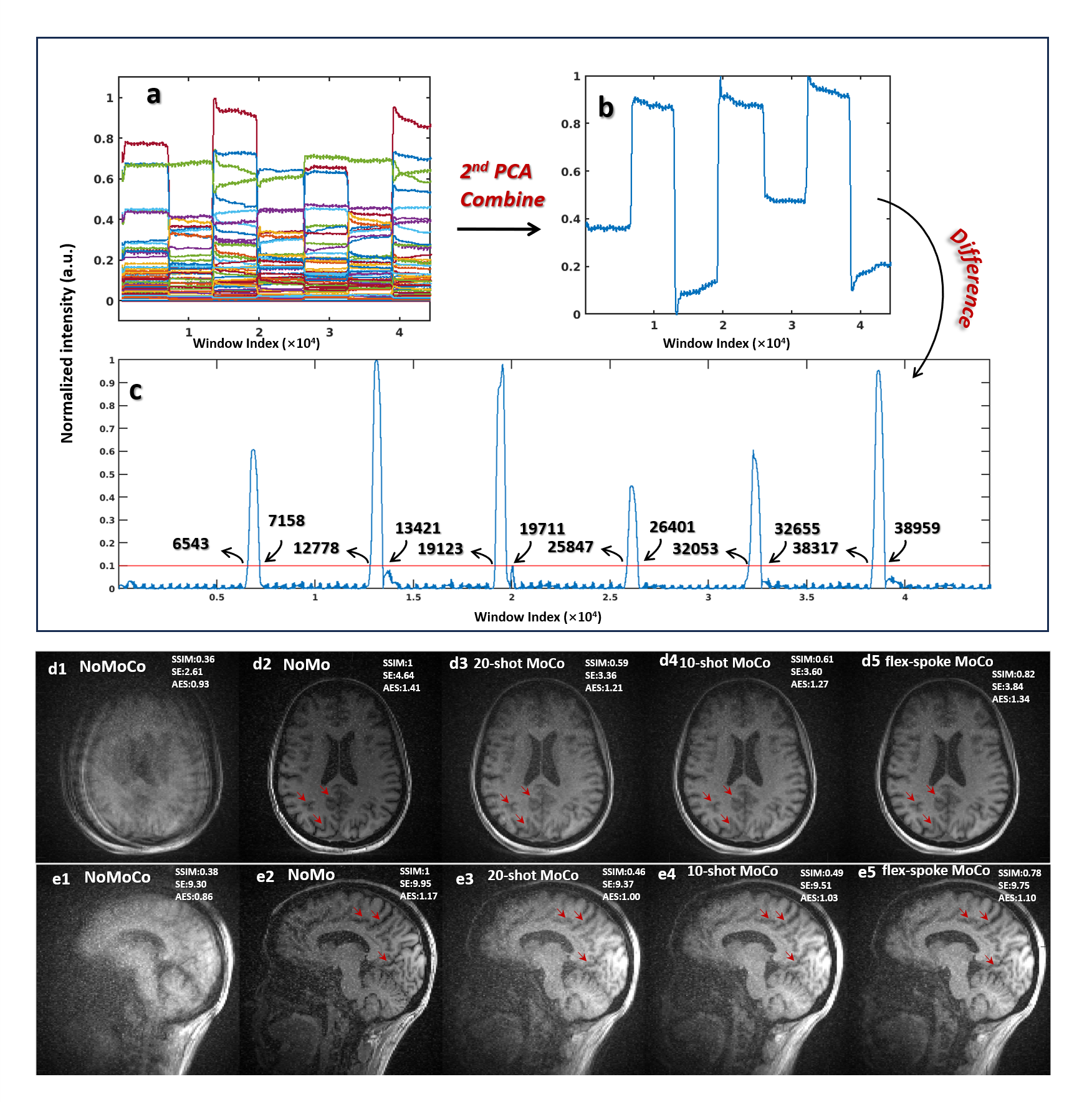}
%% Use \caption command for figure caption and label.
\caption{Results of head motion sensing and registration in Head Data 1. (a) Sliding window-summed spoke energy plot with a window length of 448 spokes, where each coil is represented by a differently colored line. (b) Normalized 2ndPCA combined signal derived from all coil signals in panel (a), providing an intuitive visualization of motion. (c) Difference of the normalized 2ndPCA combined signal, used for motion detection. Peaks in the plot indicate severe motion events, dividing the data into 7 stationary states separated by 6 peaks. The signal intensity has been normalized, and a threshold of 0.1 is applied to retrieve the starting and ending indices of the peak bandwidths as [6543, 12778, 19123, 25847, 32053, 38317] and [7158, 13421, 19711, 26401, 32655, 38959], respectively. Using these indices, 7 spoke subsets corresponding to the 7 stationary states are identified, with starting indices [1, 7157, 13420, 19710, 26400, 32654, 38958] and ending indices [6991, 13226, 19571, 26295, 32501, 38765, 44800]. These subsets are used for subsequent registration-based flexible (flex)-spoke-length motion correction, while spokes outside the 7 subsets are discarded due to severe motion corruption. (d1–d5) Reconstructed axial images from motion-corrupted data without correction (NoMoCo), no-motion data (NoMo), motion-corrected data using 20-shot Motion Correction (20-shot MoCo), 10-shot Motion Correction (10-shot MoCo), and flexible-spoke-length Motion Correction (flex-spoke MoCo). (e1–e5) Reconstructed sagittal images following the same sequence as in (d1–d5). In both (d1–d5) and (e1–e5), red arrows highlight specific anatomical structures to aid visual comparison. Additionally, Spectral Entropy (SE), Average Edge Strength (AES), and Structural Similarity Index Measure (SSIM) values are provided for quantitative evaluation. } \label{fig_H1} 
\end{figure*}

Figure \ref{fig_H1} illustrates motion sensing in head motion data from one volunteer using spoke energy and presents the results after motion correction.  The spoke energy computation for motion track in this data is under 1 sec. In Figure \ref{fig_H1}(a), the window-summed spoke energy from each coil is displayed as curves in different colors along the window index. By combining these signals into a single one using 2ndPCA, the motion trajectory is more clearly visualized, as shown in Figure \ref{fig_H1}(b). A threshold of 0.1 is applied to determine the starting and ending indices of peak bandwidths, identified as [6543, 12778, 19123, 25847, 32053, 38317] and [7158, 13421, 19711, 26401, 32655, 38959], respectively, in Figure \ref{fig_H1}(c). Using these indices and Proposition \ref{prop1}, seven spoke subsets corresponding to seven stationary states are recognized, with starting indices [1, 7157, 13420, 19710, 26400, 32654, 38958] and ending indices [6991, 13226, 19571, 26295, 32501, 38765, 44800]. These subsets are used to reconstruct images for registration-based flexible (flex)-spoke length motion correction. By employing flex-spoke subsets for registration, the quality of motion-corrected images is expected to surpass that of fixed-length shot registration. Panels d1–d5 (e1–e5) show reconstructed axial (sagittal) images from motion-corrupted data with no motion correction (NoMoCo), no motion (NoMo) data, motion-corrected data using 20-shot registration (20-shot MoCo), 10-shot registration (10-shot MoCo), and flex-spoke registration (flex-spoke MoCo). As seen in panels d1–d5 and e1–e5, the 20-shot MoCo images appear relatively blurred compared to the NoMo images, while the 10-shot MoCo images yield sharper boundaries, particularly between gray and white matter. The flex-spoke MoCo images further enhance quality, offering the finest details and closely resembling the NoMo images, particularly in anatomical regions highlighted by red arrows. The SSIM, SE, and AES values provided in the reconstructed images validate that the spoke-energy-derived flex-spoke MoCo outperforms fixed-length shot MoCo.

\begin{figure*}%% placement specifier
\centering%% For centre alignment of image.
\includegraphics[width=\linewidth]{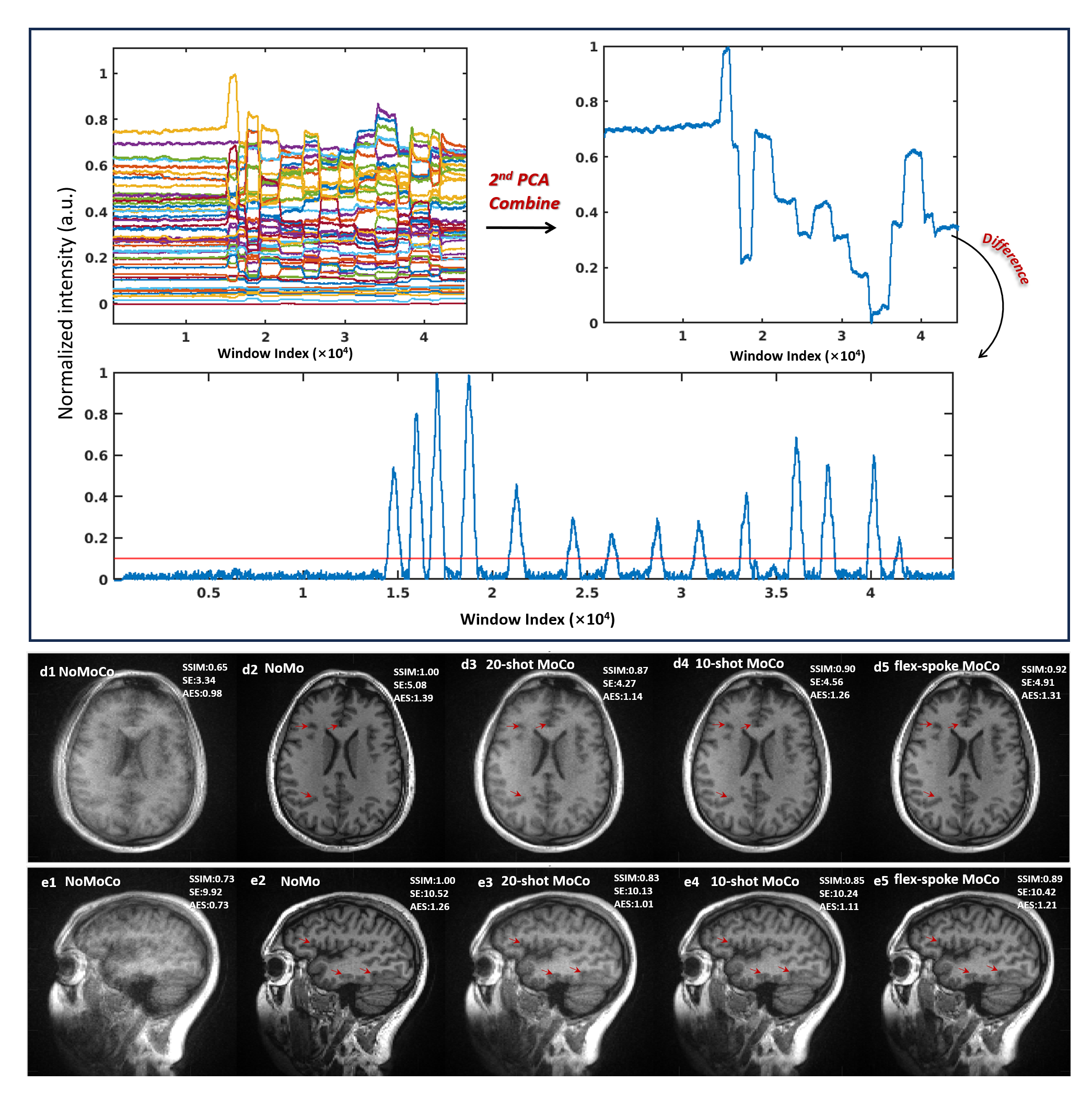}
%% Use \caption command for figure caption and label.
\caption{Results of head motion sensing and registration in Head Data 2. (a) Sliding window-summed spoke energy plot with a window length of 448 spokes, where each coil is represented by a differently colored line. (b) Normalized 2ndPCA combined signal derived from all coil signals in panel (a), providing an intuitive visualization of motion. (c) Difference of the normalized 2ndPCA combined signal, used for motion detection. Peaks in the plot indicate severe motion events, separating the data into 14 stationary states divided by 13 peaks. The signal intensity has been normalized, and a threshold of 0.1 is applied to identify 14 spoke subsets corresponding to stationary states according to \ref{prop1}, with starting indices [1, 14766, 15994, 17091, 18724, 21198, 24251, 26271, 28748, 30778, 33364, 36228, 37712, 41426] and ending indices [14767, 15995, 17092, 18725, 21199, 24252, 26272, 28749, 30779, 33365, 36229, 37713, 41427, 44800]. These subsets are used for subsequent registration-based flexible (flex)-spoke-length motion correction, while spokes outside the 14 subsets are discarded due to severe motion corruption. (d1–d5) Reconstructed axial images from motion-corrupted data without correction (NoMoCo), no-motion data (NoMo), motion-corrected data using 20-shot Motion Correction (20-shot MoCo), 10-shot Motion Correction (10-shot MoCo), and flexible-spoke-length Motion Correction (flex-spoke MoCo). (e1–e5) Reconstructed sagittal images following the same sequence as in (d1–d5). In both (d1–d5) and (e1–e5), red arrows highlight specific anatomical structures to aid visual comparison. Additionally, Spectral Entropy (SE), Average Edge Strength (AES), and Structural Similarity Index Measure (SSIM) values are provided for quantitative evaluation.} \label{fig_H2} 
\end{figure*}

Figure \ref{fig_H2} depicts similar motion sensing results from a different volunteer’s head motion data, which includes more motion events. The spoke energy computation for motion track in this data is under 1 sec. The window-summed spoke energy, shown in Figures \ref{fig_H2}(a) and (b), identifies 13 motion events. A threshold of 0.1 is applied in Figure \ref{fig_H2}(c) to define 14 spoke subsets corresponding to 14 stationary states, with starting indices [1, 14766, 15994, 17091, 18724, 21198, 24251, 26271, 28748, 30778, 33364, 36228, 37712, 41426] and ending indices [14767, 15995, 17092, 18725, 21199, 24252, 26272, 28749, 30779, 33365, 36229, 37713, 41427, 44800]. These subsets are used for registration-based flex-spoke MoCo, resulting in higher-quality motion-corrected images compared to fixed-shot MoCo. Panels d1–d5 (e1–e5) show reconstructed axial (sagittal) images from motion-corrupted data with no motion correction (NoMoCo), no motion (NoMo) data, and motion-corrected data using 20-shot (20-shot MoCo), 10-shot (10-shot MoCo), and flex-spoke registration (flex-spoke MoCo). As seen in Figure \ref{fig_H2}, 20-shot MoCo images exhibit blurring, while 10-shot MoCo images produce clearer boundaries. The flex-spoke MoCo images further enhance detail, achieving the sharpest structures and closest resemblance to the NoMo images, particularly in the anatomical regions highlighted by red arrows. Quantitative analyses using SSIM, SE, and AES values in the reconstructed images corroborate the visual assessment. 

Both experiments confirm that spoke energy is a reliable metric for motion sensing during real-time scanning. Its computational efficiency allows for on-the-fly calculation, enabling real-time motion sensing without the need for additional devices or navigator data. This approach benefits MRI practitioners by providing critical information to disregard and re-scan shots with excessive motion, supporting retrospective motion correction efforts as well.

\subsection{Stepwise Ankle Motion Tracking} 

Figure \ref{fig_A} shows the window-summed spoke energy of ankle data across 18 coils, with each color representing a distinct coil. Panel (a) displays the reconstructed image from all spokes. Panel (b) shows the window-summed spoke energy for each coil. Panel (c) presents a combined signal obtained by applying 2ndPCA to the signals in panel (b), providing a clearer and more intuitive motion trajectory. Motion events are readily reflected as peaks in panel (d), which is derived by calculating the difference of the combined signal. A threshold of 0.1 is applied to identify the starting and ending indices of four peak bandwidths as [7029, 16476, 25864, 35368] and [9792, 19068, 27275, 37732], respectively. These indices are then used to determine the spoke indices for each motion state, with starting and ending indices as [1, 9791, 19067, 27274, 37731] and [7477, 16924, 26312, 35816, 44800], obtained using Proposition \ref{prop1}, respectively.

Panels (c1–c5) display reconstructed images from the five identified spoke subsets, showcasing distinct ankle poses in each frame. This visual representation confirms the relationship between the energy peaks and the observed motion states, validating the effectiveness of spoke energy as a reliable indicator of motion during the scanning process.

\subsection{Continuous Ankle Motion Improved Reconstruction} 

Stepwise motion, commonly used in dynamic MRI protocols, requires patients to repeatedly hold still at discrete joint angles. This approach is often time-consuming, less tolerable for patients, and difficult to execute consistently, especially in pediatric or elderly populations. Additionally, stepwise schemes may suffer from uneven temporal sampling and inefficient scan coverage across the full range of motion.

In contrast, continuous cyclic motion enables subjects to move more naturally and comfortably during acquisition, improving compliance and allowing for more uniform data sampling across motion states. This acquisition mode is particularly well-suited for 4D imaging, which aims to capture smooth, temporally resolved joint dynamics. However, continuous motion introduces new challenges in reconstruction due to phase inconsistencies and spoke ordering unrelated to anatomical position.

To address this, we leverage spoke energy as a motion surrogate to reorder the k-space spokes from time-based acquisition into position-based ordering. By applying a sliding-window reconstruction on the position-reordered data, we synthesize high-quality 4D motion-resolved images with reduced artifacts and enhanced structural clarity.

We validated this strategy in two continuous motion experiments: ankle plantarflexion and knee flexion-extension.  Figures \ref{fig_CA} and \ref{fig_CK} illustrate the spoke energy–based reconstruction pipeline and comparative results for the ankle and knee continuous motion experiments. In both figures, panel (a) shows the sliding window–summed spoke energy signals across all coils, each curve representing one coil. Panel (b) presents the second principal component (2ndPCA) of these signals, which effectively captures the motion-induced variation across coils and serves as a scalar-valued positional proxy. Panel (c) displays the same signal as in (b) after sorting the spokes in ascending order of energy values, transforming the temporal acquisition into a position-representative domain.

Panels (d1–d6) show six selected frames from the 4D reconstruction using a sliding window applied to the temporally acquired spoke sequence. These frames correspond to a single slice in the coronal plane and demonstrate substantial motion blur and ghosting, due to the inconsistency of k-space sampling during continuous motion. In contrast, panels (e1–e6) show six frames reconstructed from position-ordered spokes, revealing markedly improved spatial sharpness, anatomical clarity, and consistency across time frames. Notably, joint boundaries, cartilage interfaces, and muscle contours become more distinguishable, which is critical for clinical evaluation of joint biomechanics. The related 3-planar dynamic images (cine) reconstructed using original time-ordered and position-reordered spoke data are provided in Supplementary Material.  

Together, these results confirm that the proposed spoke energy–driven reordering framework is capable of mitigating motion-induced inconsistencies and synthesizing high-quality motion-resolved 4D images from continuous acquisitions. This capability addresses a key limitation of the plain sliding-window reconstruction, offering a practical and patient-friendly solution for high-fidelity musculoskeletal motion imaging.

\subsection{PCA on Signal Combination}\label{sec_PCA}

\begin{figure*}%% placement specifier
%% Use \includegraphics command to insert graphic files. Place graphics files in 
%% working directory.
\centering%% For centre alignment of image.
\includegraphics[width=\linewidth]{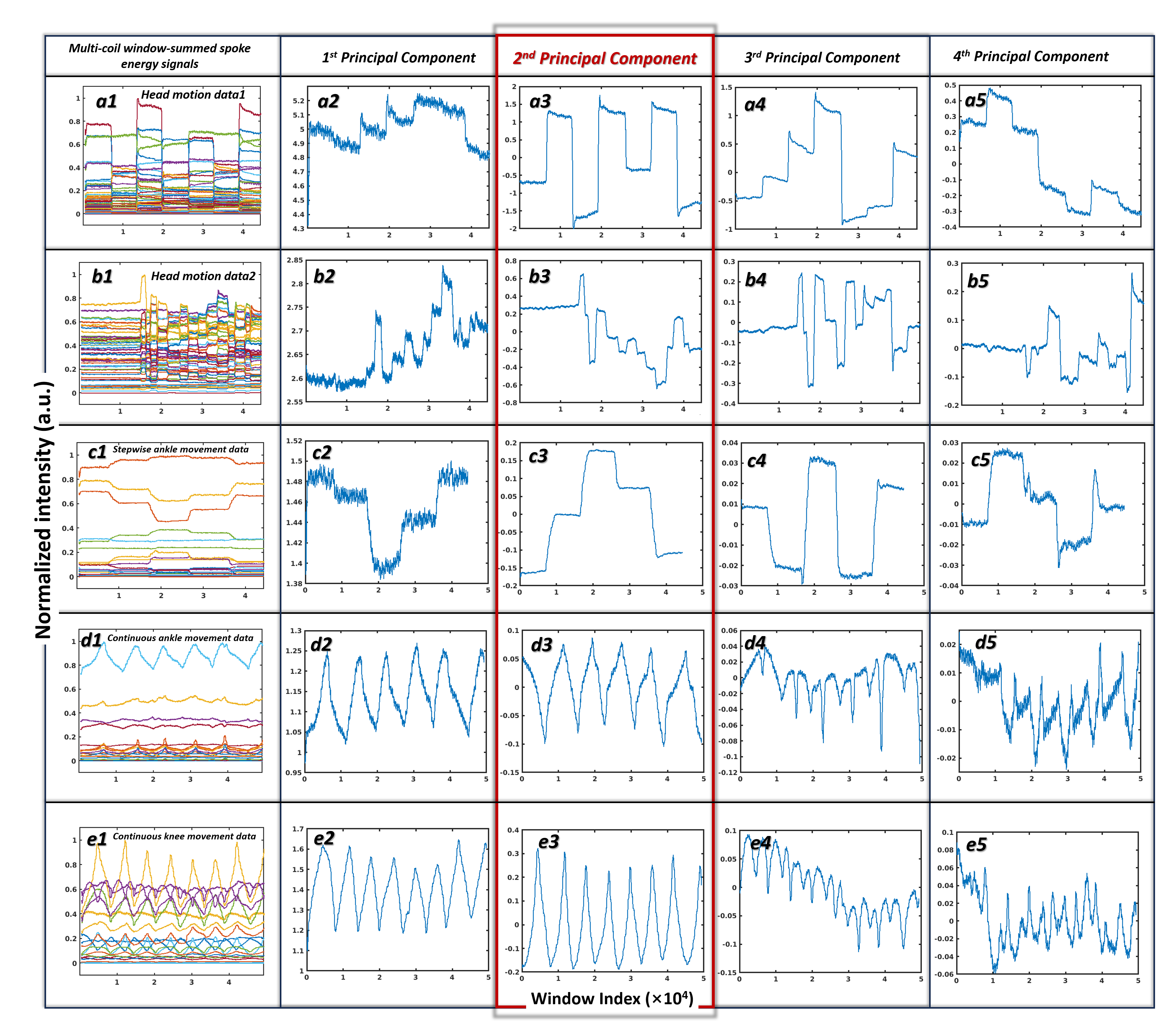}
%% Use \caption command for figure caption and label.
\caption{Principal Component Analysis (PCA) of multi-coil window-summed spoke energy signals across five representative datasets. Panels (a1–e1) display the normalized window-summed spoke energy signals from all coils for five different datasets: head motion data 1 (a1), head motion data 2 (b1), stepwise ankle movement (c1), continuous ankle movement (d1), and continuous knee movement (e1). Each colored curve corresponds to a different coil. Panels (a2–e2), (a3–e3), (a4–e4), and (a5–e5) show the 1st to 4th principal components, respectively, derived via PCA from the corresponding multi-coil signals. Notably, the 2nd principal component (highlighted in red, a3–e3) consistently captures motion-induced variations more effectively than other components. }\label{fig_PCA}
\end{figure*}

To evaluate the suitability of different PCA components for motion sensing, we performed PCA on the multi-coil window-summed spoke energy signals across five datasets used in this work. As shown in Figure \ref{fig_PCA}, the first principal components (Panels a2–e2) tend to avoid dramatic variability, even in the presence of strong motion events. This is likely because the first component captures the dominant signal intensity across coils and tends to preserve the overall signal strength. The large fluctuations in this component would compromise its energy-preserving role. Thus, high-frequency jitters may be introduced to suppress large and abrupt transitions in the signal (e.g., Panels a2–c2), which impacts motion tracking accuracy. Additionally, when some coils have significantly stronger signals than others, the first component tends to overweight those dominant coils, reducing its sensitivity to motion information present in weaker coil channels.

In contrast, the second principal component (Panels a3–e3) more reliably reflects motion-induced variations across all datasets. It avoids the saturation effects of the first component while still maintaining sufficient SNR to robustly encode motion dynamics. This makes it a more balanced and representative combination of coil information, especially when motion affects the overall signal pattern rather than just dominant coils.

Although the third principal component (Panels a4–e4) occasionally captures motion features well (e.g., Panels a4–c4), its performance is inconsistent. In lower-SNR scenarios or complex motion patterns, this component often lacks interpretability and may introduce noise. The fourth component (a5–e5) further deteriorates in quality, dominated by noise and lacking discernible motion signatures. These observations justify the selection of the second principal component as the optimal representation for multi-coil spoke energy signals in motion sensing tasks.

\subsection{Limitations and Future Directions} 

Currently, the proposed method is limited to detecting motion events and does not estimate rigid-body motion parameters, which are essential for quantitative applications such as head motion correction. Future work may explore the relationship between spoke energy signals and rigid-body motion parameters. One common strategy is to inject synthetic motion into reference image data to generate supervised training pairs of motion parameters and corresponding spoke energy signals. However, in real MPnRAGE acquisitions, spoke data are also affected by relaxation effects and other confounding factors. Therefore, accurately modeling the combined influence of these physiological and acquisition-related effects on spoke energy will be essential for reliable motion parameter estimation. Although this presents a considerable challenge, it is a worthwhile direction for advancing the clinical utility of spoke energy–based motion sensing.

%% =============================== Conclusion ============================= 

\section{Conclusions}\label{sec_conclusion}

This study presents spoke energy as a novel, self-navigated, and computationally efficient metric for real-time motion sensing in 3D radial MRI. Unlike external tracking systems or navigator-based techniques that require additional hardware, sequence modifications, or longer scan durations, the proposed approach derives motion-related information directly from k-space data, leveraging the Fourier Slice Theorem and Parseval’s Theorem. By computing window-summed spoke energy, the method captures motion-induced variations in signal intensity related to coil sensitivities, offering robust detection while minimizing the impact of noise and physiological effects. The 2ndPCA signal combination further improves interpretability, allowing for intuitive tracking of motion with a scalar-valued signal. Its simplicity and low computational burden make it well-suited for real-time feedback and seamless integration into clinical workflows.

Beyond motion detection, this work introduces a novel application of spoke energy in dynamic musculoskeletal imaging with continuous movement pattern. In particular, we demonstrate that spoke energy can serve as a proxy for object position in cyclic joint motion. By sorting the acquired spokes from temporal to position order, our method enables motion-resolved sliding-window reconstruction, improving spatial consistency and reducing motion artifacts. This framework is validated in continuous 4D imaging of the ankle and knee, where traditional stepwise acquisition protocols are less efficient and patient-friendly. The ability to generate high-quality dynamic volumes without increasing scan complexity or requiring precise timing offers new potential for assessing joint function and pathology in a clinically practical manner.

\section*{Acknowledgments}
This work was supported in part by the NIH under award numbers R01 NS124212, R01 EB019483, and R01 LM013608.

% \section*{References}

% Please ensure that every reference cited in the text is also present in
% the reference list (and vice versa).

% \section*{\itshape Reference style}

% Text: All citations in the text should refer to:
% \begin{enumerate}
% \item Single author: the author's name (without initials, unless there
% is ambiguity) and the year of publication;
% \item Two authors: both authors' names and the year of publication;
% \item Three or more authors: first author's name followed by `et al.'
% and the year of publication.
% \end{enumerate}
% Citations may be made directly (or parenthetically). Groups of
% references should be listed first alphabetically, then chronologically.

%%Harvard
% \bibliographystyle{model2-names.bst}\biboptions{authoryear}
\bibliographystyle{unsrtnat}
\bibliography{MeDIA_SpokeEnergy}

\section*{Supplementary Material}

The 3-planar dynamic images (cine) of ankle and knee reconstructed using original time-ordered and position-reordered spoke data are provided in the attached mp4 file.

% ===================  Figures ==============================================

% =========================================================================

% ========================================================================

% ========================================================================

% ========================================================================
% ========================================================================

% ========================================================================

% ========================================================================
% ========================================================================

%=======================================================================
%=======================================================================
%=======================================================================
\begin{figure*}%% placement specifier
\centering%% For centre alignment of image.
\includegraphics[width=\linewidth]{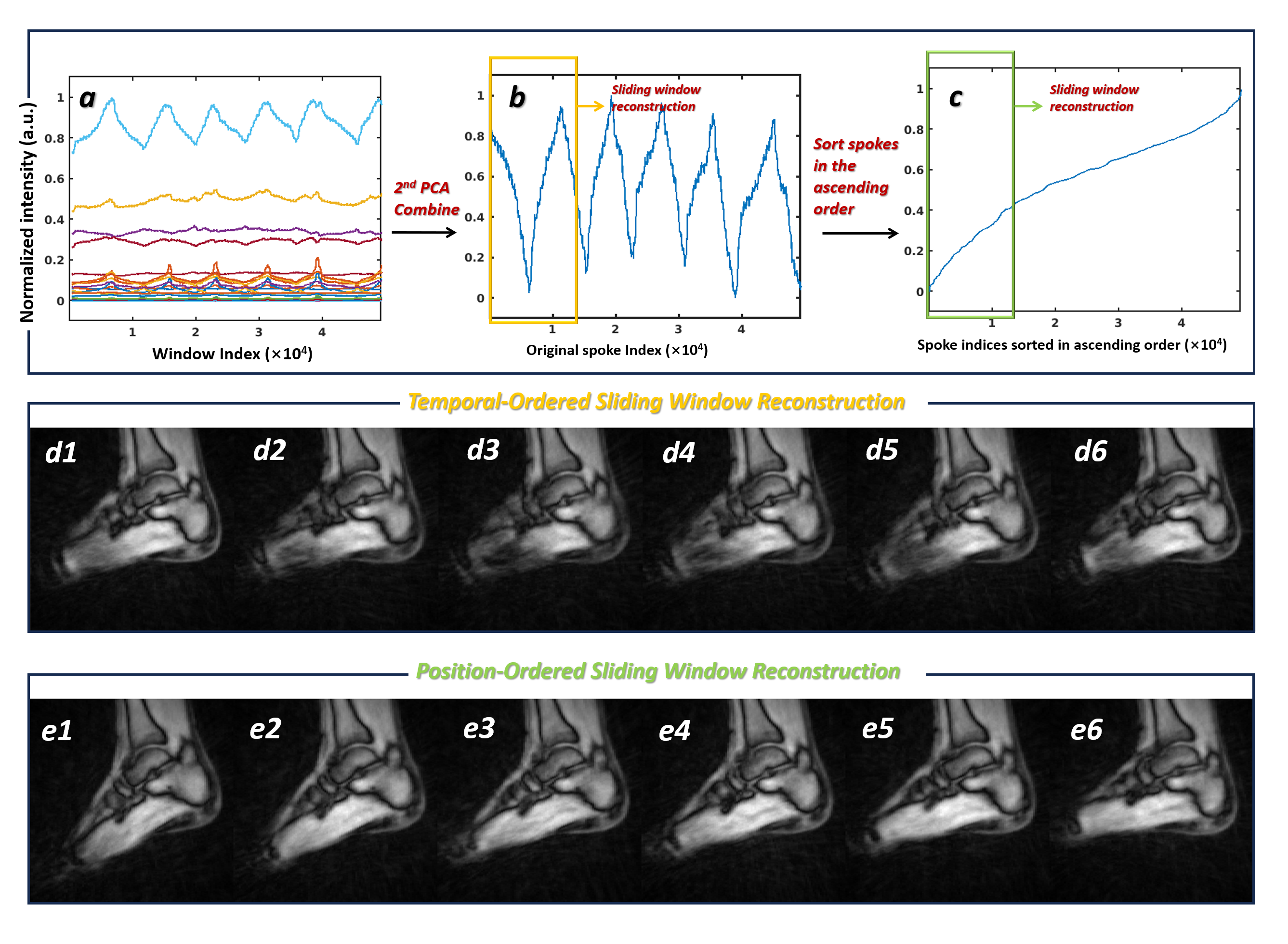}
%% Use \caption command for figure caption and label.
\caption{Analysis of continuously moving ankle data using spoke energy.(a) Sliding window–summed spoke energy signals from all coils, each represented in a different color. (b) Normalized combined signal generated by the second principal component (2ndPCA) from panel (a), intuitively tracking motion over time. (c) Same 2ndPCA signal as in (b), but with spoke indices sorted in ascending order of signal value, approximating object position rather than acquisition time. (d1–d6) Temporal-ordered sliding window reconstructions using 4480 spokes per window and a stride of 1000 spokes. Six coronal slices are selected from the resulting 4D video, showing substantial motion blur due to continuous motion. (e1–e6) Position-ordered sliding window reconstructions using the same window parameters as in (d), but with spoke data reordered by spoke energy. The resulting frames—also six coronal slices—exhibit sharper anatomical structures and more coherent motion representation.} \label{fig_CA} 
\end{figure*}

%=======================================================================
%=======================================================================

\begin{figure*}%% placement specifier
\centering%% For centre alignment of image.
\includegraphics[width=\linewidth]{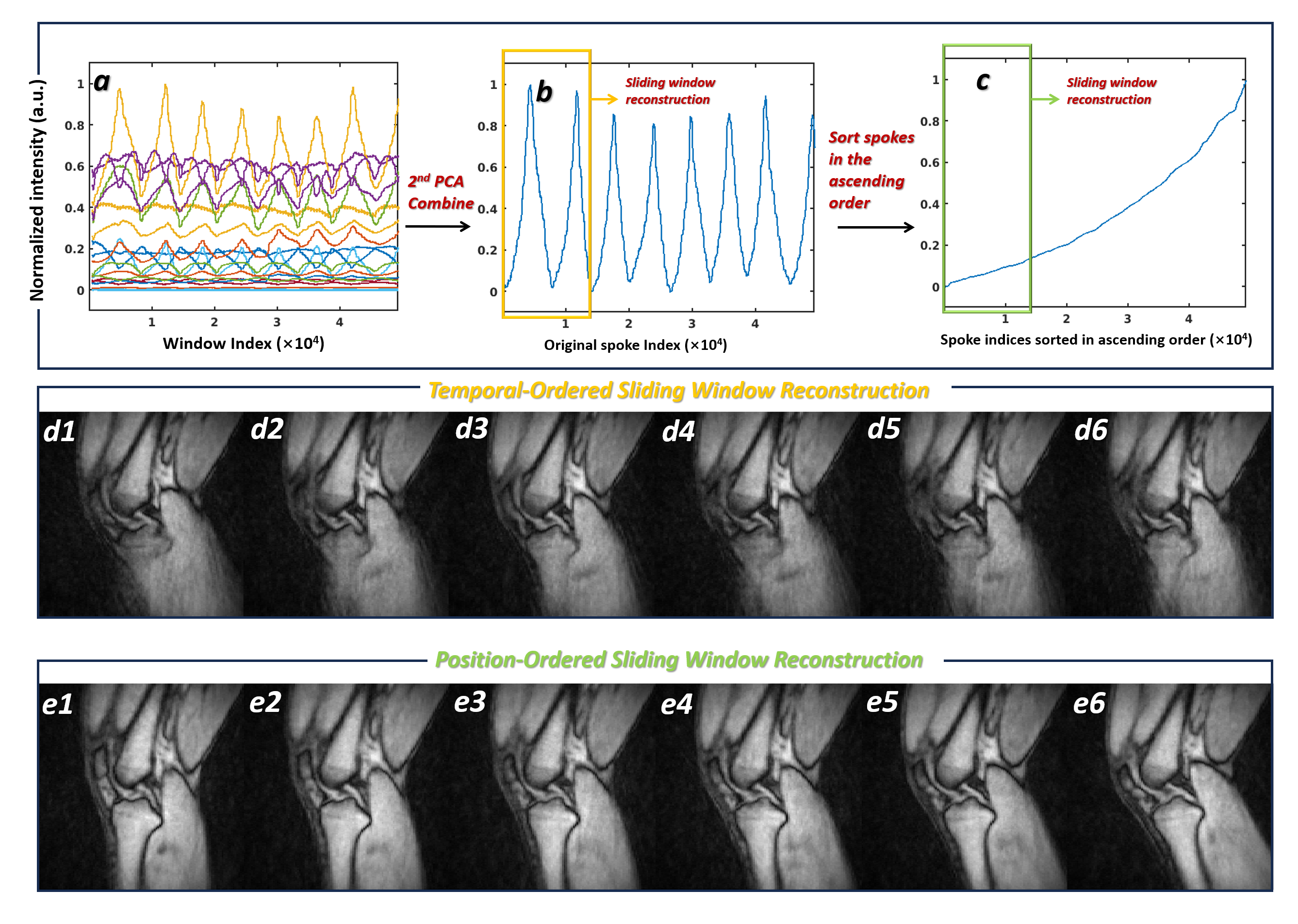}
%% Use \caption command for figure caption and label.
\caption{Analysis of continuously moving knee data using spoke energy.
Panels (a)–(e6) follow the same structure and visualization format as in the ankle experiment (Figure \ref{fig_CA}), demonstrating similar improvements in image quality and temporal consistency after spoke-energy–based spoke reordering. } \label{fig_CK} 
\end{figure*}
%=======================================================================
%=========================================================================

%=========================================================================

\end{document}